\documentclass[conference]{IEEEtran}
\IEEEoverridecommandlockouts
\usepackage{cite}
\usepackage{amsmath,amssymb,amsfonts}
\usepackage{algorithmic}
\usepackage{graphicx}
\usepackage{array}
\usepackage{makecell}
\usepackage{textcomp}
\usepackage{xcolor}
\usepackage{url}
\newcommand{\cmark}{\checkmark}
\newcommand{\xmark}{$\times$}
\def\BibTeX{{\rm B\kern-.05em{\sc i\kern-.025em b}\kern-.08em
    T\kern-.1667em\lower.7ex\hbox{E}\kern-.125emX}}
\begin{document}

\title{Academic Text-to-Music Grand Challenge: Datasets, Baselines, and Evaluation Methods}

\author{
    \IEEEauthorblockN{Fang-Chih Hsieh\textsuperscript{1*}, Wei-Jaw Lee\textsuperscript{1}, Chun-Ping Wang\textsuperscript{1}, Hung-yi Lee\textsuperscript{1}, Hao-Wen Dong\textsuperscript{2}, and Yi-Hsuan Yang\textsuperscript{1}}
    \IEEEauthorblockA{\textsuperscript{1}Artificial Intelligence Center of Research Excellence, National Taiwan University, Taipei, Taiwan}
    \IEEEauthorblockA{\textsuperscript{2}Department of Performing Arts Technology, University of Michigan, Ann Arbor, MI, United States}
    \thanks{*Corresponding author: andrew891221@gmail.com}
}
% Graduate Institute of Communication Engineering, National Taiwan University, Taipei, Taiwan

\maketitle

\begin{abstract}
This paper presents an overview and the technical framework of the ICME 2026 Grand Challenge on Academic Text-to-Music Generation (ATTM). Despite the rapid progress in text-to-music generation (TTM) systems, the field is currently dominated by models trained on massive proprietary datasets with industrial-scale computational resources, creating a significant barrier for academic research. To address this, the ATTM Challenge establishes a fair-play benchmark that requires participants to train generative models strictly from scratch using a standardized, CC-licensed subset of the MTG-Jamendo dataset containing only instrumental music. The challenge is divided into two tracks: the Efficiency Track (limited to 500M parameters) and the Performance Track (no parameter limit). Submissions are evaluated through a multi-stage process involving objective metrics, including Fréchet Audio Distance, CLAP score, and a novel Concept Coverage Score (CCS), followed by a subjective listening test. By providing open-source baselines, preprocessing pipelines, reference captions, and public evaluation code for computing FAD and CLAP, this challenge aims to facilitate and promote TTM research in academic contexts.
\end{abstract}

\begin{IEEEkeywords}
Text-to-music generation, generative AI, openness, reproducibility, affordability, large audio-language models
\end{IEEEkeywords}

\section{Introduction}
\label{sec:intro}

The landscape of generative audio has shifted toward text-to-music (TTM) systems, which use architectures such as latent diffusion and Transformers to synthesize high-fidelity music from natural language. These models offer transformative potential for creators and educators, yet progress is increasingly stalled by a compute and data wall. State-of-the-art (SOTA) systems typically rely on massive, proprietary datasets and industrial-scale hardware, creating a significant barrier for the broader academic community.  
%The landscape of generative audio has witnessed a paradigm shift in recent years, with text-to-music (TTM) generation emerging as a cornerstone of creative artificial intelligence. Powered by advances in latent diffusion models and large-scale Transformers, modern TTM systems are now capable of synthesizing high-fidelity, multi-instrument music from complex natural language descriptions. These technologies hold transformative potential for the creative industries, offering new tools for composers, content creators, and educators alike.
%However, this rapid progress is increasingly gated behind a ``compute and data wall.'' State-of-the-art models often rely on massive, proprietary datasets comprising hundreds of thousands of hours of copyrighted music and require industry-scale computational resources for training. For the broader academic community, this  creates a significant barrier for entry. 
Researchers in academic labs often find themselves limited to fine-tuning existing models or performing small-scale experiments. %that hardly generalize. 
This lack of access to standardized training conditions hampers the transparency, reproducibility, and fundamental architectural innovation necessary for the field to mature. 
Lee et al.~\cite{lee2026trainingefficienttexttomusicgenerationstatespace} investigated this question by training a few TTM models from sctratch under constrained data and computational resource. %but more work is needed.
%Lee et al.~\cite{lee2026trainingefficienttexttomusicgenerationstatespace} investigate this question by comparing the capabilities of SSM-based language models and Transformers under constrained data and computational resources. 
It motivates us to more comprehensively explore model capabilities in academic TTM settings.

%To address these challenges, 
In this paper, we present the ICME 2026 Grand Challenge on Academic Text-to-Music (ATTM) Generation.\footnote{\url{https://ntu-musicailab.github.io/ICME26-ATTM-Grand-Challenge/}} The ATTM challenge is designed as a ``fair-play'' benchmark to foster innovation under transparent and reproducible conditions. The core principle of this challenge is the requirement that all generative models must be trained from scratch using a standardized, CC-licensed dataset of 3,777 hours derived from the MTG-Jamendo corpus \cite{bogdanov2019mtg}. By restricting the data source and prohibiting the use of synthetic data generated by commercial engines, the challenge shifts the focus away from data scale and toward algorithmic efficiency, musical intelligence, and effective representation learning.

The challenge is structured into two distinct tracks to accommodate different %research goals and 
resource levels. The Efficiency Track imposes a strict limit of 500M parameters on the core generative model, encouraging the development of lightweight yet powerful architectures.
% suitable for edge deployment or academic environments. 
The Performance Track offers no parameter limits, challenging participants to push the upper bounds of musical quality within the provided data constraints. In both tracks, the task is restricted to generating 10-second instrumental music clips, which standardizes the evaluation target and keeps the benchmark practical for academic training and testing. To ensure a holistic assessment, we employ a multi-stage evaluation pipeline. Submissions are first screened using objective metrics, including Fréchet Audio Distance (FAD) \cite{kilgour2019frechetaudiodistancemetric}, CLAP scores \cite{elizalde2023clap, wu2024largescalecontrastivelanguageaudiopretraining}, and a novel Concept Coverage Score (CCS) that utilizes large audio language models (LALMs) to verify the presence of specific musical attributes. Then, top-performing systems undergo a formal mean opinion score (MOS) study conducted by listeners to evaluate musicality and prompt adherence, and to determine the final ranking.

The contributions of the challenge are four-fold:
%In this overview paper, we provide a comprehensive technical report on the ATTM Grand Challenge. Our :

\begin{enumerate}
    \item We establish a standardized fair-play benchmarking framework for TTM, providing a curated 3,777-hour dataset of instrumental music alongside automated vocal separation and captioning pipelines to ensure research transparency and reproducibility.

    \item We introduce Concept Coverage Score (CCS), a novel evaluation methodology that uses LALMs to provide a fine-grained, interpretable assessment of semantic alignment between musical concepts and generated audio.

    \item We curate in-distribution (ID) and out-of-distribution (OOD) prompt sets     featuring seen and unseen tag combinations
    for a structured evaluation benchmark, enabling systematic analysis of model performance on compositional generalizability.

    \item We provide the open-source FluxAudio \cite{fluxmusic} baseline system and a suite of training scripts, fostering accessibility and lowering the entry barrier for academic teams.
\end{enumerate}

%The remainder of this paper is organized as follows. 
%Section II defines the challenge tasks and track specifications. Section III details the dataset and preprocessing pipeline. Section IV elaborates on the evaluation methodology, and Section V presents the baseline implementation and preliminary results.

\section{Task Definition and Track Specifications}
\label{sec:task}

%The participants have to develop TTM systems under constraints. 
%Operating within a defined fair-play framework, we require that core generative models of the participants be trained entirely from scratch using a standardized public dataset.
%, thereby shifting the research emphasis from data scale and proprietary pre-training to algorithmic innovation and musical intelligence. 
%We present the core principles of the challenge and the specific constraints below.
%By establishing these boundaries, 
%the task challenges researchers to demonstrate how effectively a model can learn complex musical semantics without industrial-scale resources. 

\subsection{Key Principles and Constraints}
%To ensure a level playing field and prioritize algorithmic innovation over data scale, 
All participants must adhere to the following principles:

\begin{itemize}
    \item \textbf{Training from scratch}: The core generative model responsible for the TTM mapping must be trained entirely from scratch. The use of pre-trained weights for the main generation architecture is strictly prohibited.
    \item \textbf{Exclusive data usage}: Participants must exclusively use the provided subset of the MTG-Jamendo dataset \cite{bogdanov2019mtg} for all training and data augmentation. 
    \item \textbf{Prohibition of data laundering}: The use of external music datasets or synthetic audio generated by proprietary commercial models (e.g., Suno, Udio) is forbidden. Such practices are classified as data laundering and will result in immediate disqualification.
    \item \textbf{Auxiliary component policy:} While the core model must be trained from scratch, participants can use publicly available checkpoints for auxiliary components including:
    \begin{itemize}
        \item Audio tokenizers, autoencoders (e.g., EnCodec~\cite{defossez2022highfi}).
        \item LALMs for automated captioning/tagging.
        \item Vocoders or audio enhancement modules for post-processing.
    \end{itemize}
    However, proprietary or non-reproducible auxiliary models are prohibited.
    \item \textbf{No human-in-the-loop}: All generations must be fully autonomous, without any form of manual editing, human annotation, or cherry-picking of submitted samples.
\end{itemize}

\subsection{Track Specifications}
The challenge is divided into two tracks to accommodate varying research focuses and hardware availability.

\subsubsection{Efficiency Track}
This track is designed to encourage the development of compact and computationally efficient architectures. The core generative model is restricted to a maximum of 500M  parameters. This limit excludes auxiliary components such as text encoders, audio decoders, or vocoders. This track is particularly suited for student teams and labs focusing on edge-AI or resource-constrained optimization.

\subsubsection{Performance Track}
This track has no parameter limits for the generative model. It challenges participants to push the boundaries of musical quality and semantic alignment using the provided academic dataset, allowing for the exploration of large-scale architectures or complex ensemble methods.

\subsection{Definition of Core Generative Model}
For the purpose of parameter counting and training restrictions, the ``core generative model'' refers to the architecture responsible for the mapping from the conditioned text representation (or latent tokens) to the musical representation. It can be a latent diffusion model, a Transformer-based decoder, or a masked generative model. Encoders used solely for feature extraction and decoders used for final waveform reconstruction are considered as auxiliary.

\section{Dataset and Preprocessing}
\label{sec:dataset}

The challenge focuses on instrumental music generation, excluding singing voices. We present below the curation of the dataset, the vocal separation pipeline, and the dual-model captioning strategy used to provide diverse semantic labels.

\subsection{Source Dataset}
The challenge utilizes the MTG-Jamendo dataset \cite{bogdanov2019mtg}, which is an open-source music dataset built from tracks published on Jamendo under Creative Commons (CC) licenses, making it suitable for transparent and reproducible academic research. Specifically, we use the \texttt{raw\_30s} subset, which contains 55,701 tracks with duration longer than 30 seconds. An additional advantage of this dataset is its per-track high-quality ``tags'' annotated by human experts. These tags are organized into three  categories---genre, instrument, and mood/theme---comprising 226, 145, and 224 unique tags, respectively. As we focus on instrumental music generation, we require participants to transform this source data into a vocal-free corpus using a standardized preprocessing pipeline.

\subsection{Vocal Separation and Preprocessed Data Size}
We release the complete preprocessing codebase and require participants to perform vocal separation on their machine. 

The preprocessing uses the Mel-Band Roformer model \cite{wang2023mel}, a SOTA architecture for music source separation. Participants utilize the \texttt{melband-roformer-kim-vocals} checkpoint to isolate and remove vocal stems. To accommodate participants with limited time or computational resources, we provide an optional script to crop audio files to a shorter duration (30 seconds) prior to separation, which significantly reduces the total processing time. As a result, participants will obtain either the \emph{full} vocal-removed Jamendo dataset, spanning around 3,777 hours and occupying roughly 240~GB, or a \emph{partial} vocal-removed subset containing only the first 30 seconds of each track, spanning around 464 hours and substantially smaller at around 25~GB.

\subsection{Audio Captioning}
We offer two distinct reference caption sets generated by two different LALMs to provide semantic diversity. Participants may use these captions directly as a baseline training set, or utilize the provided captioning codebase to explore data augmentation strategies, or other captioning models.

\subsubsection{Pipeline A (``Qwen2-Audio direct captioning'')}  uses {Qwen2-Audio-7B-Instruct}~\cite{qwen2audio} to generate descriptive English captions focusing on genre, instrumentation, and mood. This pipeline produces holistic descriptions of the music’s atmosphere, disregarding any vocal elements.

\subsubsection{Pipeline B (``Music Flamingo with refinement'')} employs a two-stage approach to leverage the technical descriptive capabilities of the {Music Flamingo} model~\cite{ghosh2025musicflamingo}. Since the raw output from Music Flamingo tends to be overly verbose, we employ a second model, Qwen3-4B-Instruct, to refine and rephrase the initial descriptions. With this refinement, the captions remain concise, natural-sounding, and strictly instrumental in focus.

% The inclusion of both pipelines serves to encourage participants to investigate how caption diversity and LALM-based augmentation impact final  performance.

The specific prompts and instructions for both pipelines are summarized in Table~\ref{tab:prompts}. The inclusion of both pipelines serves to encourage participants to investigate how caption diversity and LALM-based augmentation impact final  performance.

\begin{table}[t]
\caption{Prompts for Reference Caption Generation}
\label{tab:prompts}
\centering
\begin{tabular}{p{0.18\linewidth} | p{0.72\linewidth}}
\hline
\textbf{Model / Role} & \textbf{System Prompt / Instruction} \\
\hline
Qwen2-Audio (Direct) & ``Provide a detailed English caption for this music piece, focusing on aspects such as genre, instruments used, overall mood, and production style. Do not mention anything about singing, lyrics, or vocal.'' \\
\hline
Music Flamingo (Initial) & ``Describe this track in less than 3 sentences, focusing on aspects such as genre, instruments used, overall mood, tempo and key.'' \\
\hline
Qwen3-4B (Refining) & ``Rephrase the following music caption to be shorter and more concise while keeping all essential information about genre, instruments, mood, and production style.
Remove any mention of vocals, lyrics, singers, voice types, or singing techniques. Also remove any mention of duration or length of the track.
The rephrased caption must describe only instrumental content and production characteristics, and should be fluent and natural-sounding.
Output only the rephrased caption, with no extra text, notes, or explanations.'' \\
\hline
\end{tabular}
\end{table}

\subsection{Resource Accessibility}
For transparency and lower entry barrier, the repositories for both the vocal separation pipeline\footnote{\url{https://github.com/ntu-musicailab/ICME26-ATTM-GC-Preprocessing}} and the music captioning pipelines\footnote{\url{https://github.com/ntu-musicailab/ICME26-ATTM-GC-ALM-captioning}}  are publicly available on GitHub.

\section{Topline and Baseline Models}
%\label{sec:results}

%This section demonstrates the feasibility of the challenge task and validates the proposed evaluation metrics. We provide a comprehensive benchmark comparing SOTA pre-trained models with a baseline model trained from scratch under the challenge's strict academic constraints.

Besides the submissions from the participants, we evaluate two categories of models: established pre-trained systems serving as the toplines, and our official challenge baseline.
%While the pre-trained systems are only for reference, we include the baseline to our Borda count ranking and expect the finalists to be at least better than the baseline.

\subsubsection{Topline Models}
To provide a performance ceiling, we evaluate several SOTA TTM systems, including \textbf{Stable Audio Open} (SAO)~\cite{evans2025stable}, the \textbf{MusicGen} family (small, medium, and large) \cite{copet2023simple}, and the \textbf{MeanAudio-Full} series (small and large)\cite{li2025meanaudio}. 
We use the official checkpoints for these models, which have been trained on massive proprietary or large-scale open datasets and serve as a reference for the current upper bound of unconstrained TTM performance.

\subsubsection{Official Baseline}
Our official baseline is \textbf{FluxAudio-S} (120M parameters), featuring a Flux-style Transformer architecture trained using the conditional flow matching (CFM) objective \cite{fluxmusic}.\footnote{According to the original paper \cite{li2025meanaudio}, MeanAudio is distilled from FluxAudio for single-step generation through consistency-based objectives. We use the FluxAudio architecture for our baseline to prioritize generative quality.} Instead of using pretrained checkpoints, we train this baseline from scratch using the same vocal-removed MTG-Jamendo dataset provided to the participants and the reference captions generated by Qwen2-Audio-7B-Instruct~\cite{qwen2audio}.
It utilizes a pre-trained EnCodec decoder~\cite{defossez2022highfi} for audio reconstruction and a T5-based text encoder~\cite{chung2024scaling} for prompt conditioning.
%Unlike the {MeanFlow} fine-tuning variant—which is designed for single-step generation through consistency-based objectives—our baseline focuses on the base FluxAudio model trained with CFM . 
%\subsection{Experimental Setup}
In our implementation, the baseline model was trained on a single {NVIDIA RTX A6000} with 48GB VRAM. The training proceeded for 200,000 steps with a batch size of 128.
%, utilizing approximately 45GB of VRAM. 
The total training time was about {2 days and 4 hours}. 
%This setup confirms that competitive results can be achieved with computational resources commonly available in academic environments.
%To demonstrate the accessibility of the challenge for academic labs, 
We shared the codebase of this baseline model at the outset of the competition, again to lower the entry barrier.\footnote{\url{https://github.com/ntu-musicailab/ICME26-ATTM-GC-FluxAudio}}

\section{Evaluation Methodology}

The evaluation  %for the ATTM Grand Challenge 
is divided into two phases. Phase 1 utilizes objective metrics to establish a quantitative scorecard for all submissions. Phase 2 uses a subjective MOS study to provide a qualitative assessment of the top-performing systems.

\subsection{Evaluation Data Curation}
\label{sec:evaluation}
To ensure a rigorous evaluation, we curate a set of 100 test prompts through a multi-stage pipeline involving tag filtering, stratified sampling, and LLM synthesis. This process ensures that the test prompts are musically plausible, semantically diverse, and distributionally consistent with the training data.

\subsubsection{Tag pool filtering}
We utilize the original tag metadata from the MTG-Jamendo \texttt{raw\_30s\_cleantags.csv} as the initial tag pool. To ensure that the evaluation focuses on well-defined and verifiable musical concepts, we filtered the tags based on the following four criteria:
\begin{enumerate}
    \item {Popularity}: Each tag must appear in at least 100 tracks within the original dataset.
    \item {LALM verifiability}: Using the Qwen3-Omni model~\cite{qwen3omni2025} (the judge for our CCS metric, see Section \ref{sec:obj}), each tag must achieve a recall rate $\ge 0.85$ based on the MTG-Jamendo ground truth.
    \item {Reference consistency}: Each tag must appear at least 10 times across the two reference caption sets we provide.
    \item {Instrumental constraint}: All vocal-related tags such as ``choir'' and ``vocals'' are manually excluded.
\end{enumerate}
This filtering leads to a final taxonomy of 143 valid tags, comprising 55 genres, 25 instruments, and 63 moods/themes.

\subsubsection{Stratified triplet sampling}
From the filtered taxonomy, we sampled 100 unique triplets of tags, each consisting of exactly one genre, one instrument, and one mood/theme tag. To assess the models' generalization capabilities while maintaining fairness, we categorized these triplets based on their co-occurrence patterns in the original dataset:
\begin{itemize}
    \item \textbf{In-Distribution} (ID): 80 triplets were sampled where all tag pairs in a triplet (i.e., ``genre \& instrument'', ``genre \& mood/theme '', and ``instrument \& mood/theme'')have co-occurred at least once in the training data, not necessarily in the same songs.
    \item \textbf{Out-of-Distribution} (OOD): 20 triplets were sampled where at least one tag pair in a triplet has never co-occurred (e.g., ``heavy metal'' \& ``calm''). 
\end{itemize}
While OOD samples are included to explore model behavior in edge cases for future work, only scores achieved on the 80 ID prompts contribute to the final official leaderboard ranking. The reason is that our objective evaluation metrics have not yet been verified on the OOD samples, and therefore may not produce sufficiently reliable scores for official ranking.

\subsubsection{Prompt synthesis via LLM}
The sampled triplets were transformed into fluent, descriptive English captions using Qwen3-4B-Instruct. To prevent distributional shift between the test prompts and the provided training captions, we employ a 10-shot in-context learning (ICL) approach, providing the model with random examples/demonstrations from the reference caption sets (5 shots from captioning pipeline A and 5 shots from pipline B) to match the tone and phrasing of the employed music captioning pipelines. 

Notably, we apply two synthesis strategies for the triplets:
\begin{enumerate}
    \item \textbf{Strict following} (40 ID and 20 OOD): The model is instructed to describe the music using exclusively the provided tags, not adding other musical elements.
    \item \textbf{Improvisation} (40 ID): To improve musical coherence and simulate realistic descriptions, the model is prompted to improvise by adding one to three additional, musically plausible instruments to the caption. This improvisation is only used to enrich the prompt text: the enriched prompt is used for computing the CLAP score, but the added instruments are not counted as target concepts in the CCS evaluation.
\end{enumerate}
The resulting 100 prompts serve as the final evaluation set, providing a balanced mix of strict tag adherence and naturally enriched musical descriptions.

\subsection{Objective Evaluation (Phase 1)}
\label{sec:obj}
In the first phase, submissions are ranked using a composite objective score. Each system is evaluated based on 10-second generated audio clips, matching the challenge task definition. While standard metrics such as FAD and CLAP scores are employed, we emphasize the use of high-correlation embedders and introduce a novel granular metric named CCS for semantic verification.
To facilitate transparent and reproducible benchmarking, we also release the official evaluation code for computing FAD and CLAP to the public.\footnote{\url{https://github.com/ntu-musicailab/ICME26-ATTM-GC-Evaluation}}

\subsubsection{\textbf{Fréchet Audio Distance}}
FAD evaluates the distributional similarity between generated audio and a reference set of real music. We adopt the LAION-CLAP-Music model \texttt{music\_audioset\_epoch\_15\_esc\_90.14} as the feature extractor, as it provides FAD scores most aligned with human preference according to \cite{biswas2025towards, grotschla2025benchmarking}. For the reference set, we use a hidden instrumental subset from MTG-Jamendo consisting of 1{,}000 randomly sampled tracks.

\subsubsection{\textbf{CLAP Score}}
To measure the global semantic alignment between the input prompt $y$ and the generated audio $\mathbf{x}$, we compute the cosine similarity in the joint embedding space of the CLAP model. Following \cite{grotschla2025benchmarking}, we employ the same \texttt{music\_audioset\_epoch\_15\_esc\_90.14} checkpoint, which exhibits the highest correlation with human ratings regarding text-to-audio relevance in the music domain.

\subsubsection{\textbf{Concept Coverage Score} (CCS)}
While the CLAP score provides a global measure of semantic similarity, it lacks interpretability regarding specific musical attributes. To address this, we propose CCS, a novel metric that utilizes the SOTA LALM, Qwen3-Omni~\cite{qwen3omni2025}, as a zero-shot music judge to verify the presence of individual musical tags. Specifically, given a test prompt synthesized from a triplet of tags $T = \{t_g, t_i, t_m\}$ representing {genre}, {instrument}, and {mood/theme} respectively, CCS assesses how many of these target concepts can be detected from the generated audio $\mathbf{x}$.

\paragraph{LALM-based verification}
For each concept $t \in T$,  LALM is provided with the audio $\mathbf{x}$ and a category-specific prompt. These prompts instruct the model to act as a specialized classifier (e.g., a music genre classifier or an audio event detector) and determine if any trace of the concept $t$ is present. To ensure robustness and bypass potential linguistic biases in text generation, we do not rely on the model's textual output. Instead, we extract the log-probabilities (logits) for the tokens ``Yes'' and ``No.'' The detection function $D(\mathbf{x}, t)$ is defined as:
\begin{equation}
D(\mathbf{x}, t) = 
\begin{cases} 
1 & \text{if } \text{logit}(\text{``Yes''}) > \text{logit}(\text{``No''}) \,,\\
0 & \text{otherwise} \,.
\end{cases}
\end{equation}

\paragraph{Scoring and aggregation}
We report a single aggregate CCS that measures the fraction of target concepts detected across all evaluation samples. Specifically, for a set of $N$ evaluation samples $\{\mathbf{x}_i, T_i\}_{i=1}^N$, the score is calculated as:
\begin{equation}
\text{CCS}
= \frac{1}{3N} \sum_{i=1}^{N} \sum_{t \in T_i} D(\mathbf{x}_i, t) \,.
\end{equation}

\begin{table*}[t]
\caption{Comprehensive benchmarking summary for the objective evaluation using the final test prompts. }
\label{tab:main_results}
\centering
\setlength{\tabcolsep}{5pt}
\renewcommand{\arraystretch}{1.15}
\footnotesize
% \resizebox{\textwidth}{!}{%
\begin{tabular}{l | r r c c c c c r | c c c | c}
\hline
\textbf{Model} & \textbf{Params} & \makecell{\textbf{Train Data}\\\textbf{(hours)}} & \textbf{Arch.} & \makecell{\textbf{Official}\\\textbf{Caption}} & \makecell{\textbf{MTG}\\\textbf{Tags}} & \makecell{\textbf{Post-}\\\textbf{train}} & \makecell{\textbf{Infer.}\\\textbf{Opt.}} & \makecell{\textbf{A100-eq}\\\textbf{GPU-H}} & \textbf{FAD}\,$\downarrow$ & \textbf{CLAP}\,$\uparrow$ & \textbf{CCS}\,$\uparrow$ & \textbf{Rank} \\
\hline
Submission \texttt{e00} \cite{assp2026meanaudio} & 120M & 0.46K & D/F,\,T & \cmark & \xmark & \xmark & \xmark & 48.0 & 0.556 & \underline{0.310} & 0.796 & 6 \\
Submission \texttt{e01} \cite{wang2026icmew} & 189M & 3.7K & D/F,\,SSM & \cmark & \xmark & \xmark & \xmark & 25.1 & 0.577 & \textbf{0.338} & \underline{0.863} &  2\\
Submission \texttt{e02} \cite{kim2026improving} & 120M & 3.7K & D/F,\,T & \cmark & \xmark & \cmark & \xmark & 14.2 & 0.498 & 0.270 & 0.763 & 8 \\
Submission \texttt{e03} & 340M & 0.46K & D/F & \cmark & \cmark & \xmark & \cmark & 127.0 & 0.518 & 0.251 & 0.763 & 12 \\
Submission \texttt{e04} & 70M & 0.46K & D/F,\,T & \cmark & \xmark & \xmark & \xmark & 3.8 & 0.574 & 0.195 & 0.833 & 9 \\
Submission \texttt{e05} \cite{koh2026instrumental} & 499M & 0.46K & D/F,\,T & \cmark & \xmark & \cmark & \cmark & 20.0 & 0.487 & 0.305 & 0.800 & 2 \\
Submission \texttt{e06} & 488M & 0.46K & T, SSM & \cmark & \xmark & \xmark & \cmark & 38.1 & 0.667 & 0.268 & 0.808 & 9 \\
Submission \texttt{e07} \cite{chen2026s2accompanist} & 402M & 3.7K & D/F & \xmark & \xmark & \cmark & \xmark & 148.9 & \textbf{0.417} & 0.261 & \textbf{0.867} & 1 \\
Submission \texttt{e08} \cite{scoreawarettm} & 450M & 3.7K & D/F,\,T & \cmark & \xmark & \xmark & \xmark & 190.3 & 0.495 & 0.295 & 0.804 & 2 \\
Submission \texttt{e09} \cite{UT-AISTimprtATTM2026} & 480M & 3.7K & D/F,\,T & \cmark & \xmark & \xmark & \xmark & 168.1 & 0.646 & 0.263 & 0.767 & 12 \\
Submission \texttt{e10} \cite{cheng2026modeling} & 315M & 0.46K & T & \cmark & \xmark & \xmark & \xmark & 482.1 & \underline{0.482} & 0.163 & 0.738 & 11 \\
Submission \texttt{e11} & 480M & 0.46K & D/F & \cmark & \xmark & \cmark & \xmark & 84.6 & 0.892 & 0.097 & 0.675 & 16 \\
\hline
Submission \texttt{p00} \cite{assp2026meanaudio} & 502M & 0.46K & D/F,\,T & \cmark & \xmark & \xmark & \xmark & 48.0 & 0.557 & \textbf{0.311} & \underline{0.796} & 6 \\
Submission \texttt{p05} \cite{koh2026instrumental} & 2.4B & 0.46K & D/F,\,T & \cmark & \xmark & \cmark & \cmark & 60.0 & \underline{0.514} & \underline{0.306} & \textbf{0.800} & 5 \\
Submission \texttt{p09} \cite{UT-AISTimprtATTM2026} & 480M & 3.7K & D/F,\,T & \cmark & \xmark & \xmark & \xmark & 168.1 & 0.646 & 0.260 & 0.767 & 15 \\
Submission \texttt{p10} \cite{cheng2026modeling} & 1.5B & 0.46K & T & \cmark & \xmark & \xmark & \xmark & 1319.3 & \textbf{0.500} & 0.171 & 0.721 & 14 \\
\hline
FluxAudio-S (Baseline) & 120M & 3.7K & D/F,\,T & \cmark & \xmark & \xmark & \xmark & 25.8 & 0.757 & 0.088 & 0.592 & 17 \\
\hline
Stable\,Audio\,Open~\cite{evans2025stable} & 1.1B & 7.3K & D/F & -- & -- & -- & -- & N/A & 0.574 & 0.321 & 0.800 & -- \\
MusicGen-small~\cite{copet2023simple} & 300M & 20K & T & -- & -- & -- & -- & N/A & 0.574 & 0.370 & 0.875 & -- \\
MusicGen-medium~\cite{copet2023simple} & 1.5B & 20K & T & -- & -- & -- & -- & N/A & 0.548 & 0.353 & 0.892 & -- \\
MusicGen-large~\cite{copet2023simple} & 3.3B & 20K & T & -- & -- & -- & -- & N/A & 0.553 & 0.379 & 0.888 & -- \\
MeanAudio-S-Full~\cite{li2025meanaudio} & 120M & 10K & D/F,\,T & -- & -- & -- & -- & N/A & 0.649 & 0.210 & 0.808 & -- \\
MeanAudio-L-Full~\cite{li2025meanaudio} & 480M & 10K & D/F,\,T & -- & -- & -- & -- & N/A & 0.660 & 0.202 & 0.783 & -- \\
\hline
\end{tabular}
% }
\\[2pt]
\parbox{\textwidth}{\footnotesize\textit{Note:} Submission letters indicate the challenge track: \texttt{e} denotes the Efficiency Track and \texttt{p} denotes the Performance Track. The ``Train Data (h)'' column reports the amount of training audio in thousands of hours; 3.7K indicates use of the full MTG-Jamendo dataset (about 3,777 hours), whereas 0.46K indicates use of the 30-second subset (approximately 464 hours). ``Arch.'' stands for architecture; ``D/F'' denotes diffusion/flow-matching, ``T'' denotes Transformer, and ``SSM`` denotes state space models. ``Official Caption'' indicates whether a system uses the two provided official caption sets; if not, the team curated an alternative caption set, for example through data augmentation. ``MTG Tags'' indicates whether the original MTG-Jamendo tags are used during training. ``Post-train'' denotes whether post-training techniques are used. ``Infer. Opt.'' denotes whether inference-time optimization techniques are used. ``A100-eq GPU-H'' denotes hardware-normalized training GPU-hours, computed as self-reported GPU-hours multiplied by the dense FP16 Tensor throughput of the reported GPU and divided by 312 TFLOP/s, the A100 80GB reference throughput; this is a proxy for comparable training cost rather than measured realized FLOPs. We highlight the best result in each track, and underscore the second best. ``Rank'' indicates the ranking of the objective result. }
\end{table*}

\subsection{Final Ranking of the Objective Result via Borda Count}
\label{sec:borda}

We employ the Borda count method to convert the scores in the three objective metrics into a single leaderboard.
Specifically, for a ranking pool with $M$ submissions, we rank all submissions plus the FluxAudio baseline (i.e., there are in total $M+1$ candidates) for each objective metric independently, in ascending order for FAD, and in descending order for CLAP and CCS. Let $r_m(s) \in \{1, \dots, M+1\}$ denote the rank of submission $s$ under metric $m \in \{\mathrm{FAD}, \mathrm{CLAP}, \mathrm{CCS}\}$, where rank $1$ is the best. The Borda score assigned to submission $s$ for metric $m$ is then
$B_m(s) = M + 1 - r_m(s)$.
% \begin{equation}
% B_m(s) = M + 1 - r_m(s) \,.
% \end{equation}
Thus, the best submission for a metric receives $M$ points (i.e., the higher the better), the second-best receives $M-1$ points, etc, until the lowest-ranked submission receives $0$ points.

The overall objective score of submission $s$ is the sum of its Borda scores across the three metrics:
\begin{equation}
B_{\mathrm{total}}(s) = B_{\mathrm{FAD}}(s) + B_{\mathrm{CLAP}}(s) + B_{\mathrm{CCS}}(s)\,.
\end{equation}
Submissions are then sorted by $B_{\mathrm{total}}(s)$ to obtain the objective ranking.
The toplines, baseline and submissions are all evaluated on the 80 ID test prompts curated in Section~\ref{sec:evaluation}.

Since each team may submit up to two results per track, we use a two-stage ranking procedure. In the first stage, we run the Borda count separately within the Efficiency Track and the Performance Track. If a team has two submissions in the same track, we retain only the higher-ranked one as that team's final entry for that track. After this filtering step, each team now has at most one submission in each track.

In the second stage, we merge the remaining submissions from both tracks into a joint submission pool and run the Borda count again. The resulting scores are used to produce the official objective ranking and to determine which teams advance as finalists to the subjective evaluation phase. We select the top 6 submissions in this ranking as the finalists, requiring that every finalist beat the baseline.

This aggregation scheme has two advantages. First, it balances the three objective criteria without requiring manual tuning of metric weights. Second, it discourages over-optimization toward any single metric, since a strong final ranking requires good performance across fidelity, global text-audio alignment, and concept-level semantic coverage.

\subsection{Objective Benchmarking Results}
The challenge was officially launched on February 10, 2026, with the final test prompts released on April 20, 2026, and the final audio submission deadline on April 23, 2026. In total, 18 teams from around the world registered, and we finally received submissions from 12 teams for the Efficiency Track, among which 4 teams also submitted to the Performance Track. Since some teams submitted two systems within a track, we retained only the higher-ranking one using the Borda count procedure described in Section~\ref{sec:borda}. Table~\ref{tab:main_results} reports the final submissions, our baseline, and the toplines, and summarizes their model size, training data, core-model architecture, design choices, and objective results.

All submissions to either track beat the FluxAudio baseline. Following the objective ranking, we select four teams from the Efficiency Track (\texttt{e01}, \texttt{e05}, \texttt{e07}, \texttt{e08}) and two teams from the Performance Track (\texttt{p00}, \texttt{p05}) as the finalists.

% Following the objective ranking, submission \texttt{e07} is ranked 1\textsuperscript{st}; \texttt{e01}, \texttt{e05}, and \texttt{e08} obtain the same second-highest Borda count and are all tied for 2\textsuperscript{nd}; \texttt{p05} is ranked 5\textsuperscript{th}; and \texttt{p00} and \texttt{e00} obtain the same Borda count and are tied for 6\textsuperscript{th}. As \texttt{p00} and \texttt{e00} are from the same research team, we select \texttt{p00} to break the tie.
% As a result, 4 teams from the Efficiency Track (\texttt{e01}, \texttt{e05}, \texttt{e07}, \texttt{e08}) and 2 teams from the Performance Track (\texttt{p00}, \texttt{p05}) form the finalists.

%for each track advance to the subjective evaluation phase, including \texttt{e01}, \texttt{e05}, \texttt{e07}
%%, \texttt{e08} 
%for the Efficiency Track and 
%\texttt{p00}, \texttt{p05}, \texttt{p10} for the Performance Track (\texttt{e05} and \texttt{p05} are from the same research team).

\subsection{Subjective Evaluation (Phase 2)}

For the listening test, we compare the 6 finalists chosen from the two tracks, along with a topline model.
Specifically, we choose MusicGen-small \cite{copet2023simple}, for it is the only MusicGen series with fewer than 500M parameters, making it compatible with our parameter limit.
We note that this official MusicGen-small checkpoint was trained on ``20K hours of licensed music'' \cite{copet2023simple}, which is at least five times larger than our MTG-Jamendo dataset.
We view MusicGen-small as a high anchor, not requiring any finalist to outperform it.

Therefore, the listening test compares in total 7 model variants, including the official MusicGen-small and the 6 finalists. Each questionnaire contains 5 prompts randomly sampled from the 100 test prompts (1 OOD and 4 ID, two of which involve improvisation), along with the 10-second clips generated by the 7 variants, yielding 35 evaluation items per questionnaire. For broader prompt coverage while keeping the listening load manageable, we form 5 independent questionnaires using different prompt subsets, so that each variant is evaluated on 25 samples in total across the full study. We distribute the questionnaires through mailing lists and ask volunteer evaluators to answer one of them, preserving matched comparisons among variants within each prompt while distributing listener effort across the questionnaires. The evaluators will rate the generated samples on a 5-point Likert scale (the higher is better) across the following 4 criteria:

\begin{itemize}
    \item \textbf{Audio Fidelity}: Evaluates the technical production quality, signal clarity, and the absence of perceptible digital artifacts or distortion.
    \item \textbf{Prompt Adherence}: Assesses how accurately the generated musical elements, such as instrumentation and genre, correspond to the attributes requested in the text prompt.
    \item \textbf{Musicality}: Measures intrinsic musical quality, including rhythmic stability, harmonic coherence, structural development, and general listenability.
    \item \textbf{Overall}: Captures the holistic impression of the generated music by integrating production quality, prompt relevance, and musical appeal into a single overall judgment.
\end{itemize}

%We will report the results of the listening test here later.
%We plan to use the Overall MOS for the final ranking.

\subsection{Subjective Evaluation Results}

We conducted the subjective listening test %from May 6 to May 12, 2026, 
and received 35 participant responses 
in total.
%across five questionnaires. 
Among these responses, 25 were provided by expert listeners. Expert listeners are defined as participants who have more than three years of musical background, currently work in a music-related profession, or rate their music appreciation level above 3. Table \ref{tab:mos_results} reports the final subjective evaluation results based on the ``Overall'' MOS, along with (`$\pm$') the standard deviations. We report two sets of MOS scores: MOS\_all, which includes responses from all participants, and MOS\_expert, which includes only responses from expert listeners.

\section{Conclusion}
The ATTM Grand Challenge establishes a rigorous foundation for academic research in text-to-music generation. By standardizing the dataset, providing a transparent evaluation protocol through the novel CCS objective evaluation metric, and offering a reproducible FluxAudio baseline, we aim to encourage algorithmic breakthroughs that are independent of massive industrial-scale computational resources. We invite the community to participate in this effort to build more intelligent, efficient, and interpretable music generation systems.

% \begin{table}[t]
% \caption{Subjective evaluation results based on overall MOS scores}
% \label{tab:mos_results}
% \centering
% \renewcommand{\arraystretch}{1.15}
% \begin{tabular}{l | c c | c}
% \hline
% \textbf{Model} & \textbf{MOS\_all} & \textbf{MOS\_expert} & \textbf{Award} \\
% \hline
% Submission \texttt{e01} & 3.225 & 3.117 & Efficiency 2nd \\
% Submission \texttt{e05} & 2.969 & 2.929 & -- \\
% Submission \texttt{e07} \cite{chen2026s2accompanist} & 3.250 & 3.186 & Efficiency 1st \\
% Submission \texttt{e08} & 3.119 & 3.044 & Efficiency 3rd \\
% \hline
% Submission \texttt{p00} \cite{assp2026meanaudio} & 2.006 & 2.044 & -- \\
% Submission \texttt{p05} & 3.344 & 3.327 & Performance 1st \\
% \hline
% MusicGen-small \cite{copet2023simple} & 3.538 & 3.425 & -- \\
% \hline
% \end{tabular}
% \end{table}

\begin{table}[t]
\caption{Subjective evaluation results based on overall MOS scores}
\label{tab:mos_results}
\centering
\renewcommand{\arraystretch}{1.15}
\resizebox{\columnwidth}{!}{%
\begin{tabular}{l | c  | c  | c}
\hline
\textbf{Model} & \textbf{MOS\_all}  & \textbf{MOS\_expert} & \textbf{Award} \\
\hline
Submission \texttt{e01} \cite{wang2026icmew} & 3.225 {\scriptsize{$\pm$\,1.093}} & 3.177 {\scriptsize $\pm$\,1.136} & Efficiency 2nd \\
Submission \texttt{e05} \cite{koh2026instrumental} & 2.969 {\scriptsize $\pm$\,1.194} & 2.929 {\scriptsize $\pm$\,1.230} & -- \\
Submission \texttt{e07} \cite{chen2026s2accompanist} & 3.250 {\scriptsize $\pm$\,1.234} & 3.186 {\scriptsize $\pm$\,1.286} & Efficiency 1st \\
Submission \texttt{e08} \cite{scoreawarettm} & 3.119 {\scriptsize $\pm$\,1.084} & 3.044 {\scriptsize $\pm$\,1.113} & Efficiency 3rd \\
\hline
Submission \texttt{p00} \cite{assp2026meanaudio} & 2.006 {\scriptsize $\pm$\,1.031} & 2.044 {\scriptsize $\pm$\,1.089} & -- \\
Submission \texttt{p05} \cite{koh2026instrumental} & 3.344 {\scriptsize $\pm$\,1.116} & 3.327 {\scriptsize $\pm$\,1.137} & Performance 1st \\
\hline
MusicGen-small \cite{copet2023simple} & 3.538 {\scriptsize $\pm$  1.009} & 3.425 {\scriptsize $\pm$  0.998} & -- \\
\hline
\end{tabular}
}
\end{table}

%\begin{figure}[t]
%    \centering
%    \includegraphics[width=\linewidth]{MOS.png}
%    \caption{Subjective evaluation results based on overall MOS scores.}
%    \label{fig:mos_results}
%\end{figure}

\section*{Acknowledgment}
This work is supported by grants from the Ministry of Education\,(MOE) of Taiwan (for Taiwan Centers of Excellence), the National Science and Technology Council of Taiwan (NSTC 114-2628-E-002-013-MY3), and Google Asia Pacific. We are also grateful for the support of Moises AI in providing cash awards for the challenge winners.

\bibliographystyle{IEEEbib}
\bibliography{icme2026references}

\end{document}